\documentclass{llncs}
\usepackage{svg}
\usepackage{xifthen}
\usepackage{pgfplots}
\usepackage{amsmath}
\usepackage{pdfpages}
\usepackage{tikz}
\usepackage{comment}
\usepackage{hyperref}
\usepackage{algorithm}
\usepackage{mathtools}
\usepackage{amssymb}
\usepackage{todonotes}
\usepackage[noend]{algpseudocode}
\usetikzlibrary{calc}
\newtheorem{observation}{Issue}
\usetikzlibrary{shapes,snakes}
\usetikzlibrary{arrows,backgrounds,snakes}
\usepackage{varwidth}
\usepackage{subcaption}

\usepackage[utf8x]{inputenc}

\newcommand{\Val}{\mathit{Val}} 
\newcommand{\Var}{\mathit{Var}} 
\newcommand{\commit}{\texttt{cmt}}
\newcommand{\abort}{\texttt{abrt}}
\newcommand{\rd}{\texttt{rd}}
\newcommand{\wrt}{\texttt{wr}}
\newcommand{\bbC}{\mathbb{C}}
\newcommand{\bbR}{\mathbb{R}}
\newcommand{\bbA}{\mathbb{A}}

\begin{document}

\title{On the (Non-)Applicability of a Small Model Theorem to Model Checking STMs\thanks{This work 
has been partially supported by DFG grant WE 2290/12-1.}}

\author{Heike Wehrheim}

\institute{ University of Oldenburg \\ Department of Computer Science \\ Germany}

\maketitle
\begin{abstract}
Software Transactional Memory (STM) algorithms provide programmers with a synchronisation mechanism 
for concurrent access to shared variables. Basically, programmers can specify {\em transactions} (reading from and writing to shared state) 
which execute  
``seemingly'' atomic. This property is captured in a correctness criterion called {\em opacity}. 
For model checking opacity of an STM algorithm, we -- in principle -- need to check opacity 
for all possible combinations of transactions writing to and reading from 
potentially unboundedly many variables. 
 
To still apply automatic model checking techniques to opacity checking, a so called 
{\em small model theorem} has been proven which states that model checking on 
two variables and two transactions is sufficient for correctness verification of 
STMs. In this paper, we take a fresh look at this small model theorem and 
investigate its applicability to opacity checking of STM algorithms. 
\end{abstract}

\section{Introduction}
\label{intro}
Today, multi-core processors are widely utilized since their usage yields a large increase in computing power. 
This additional computing power can best be employed in concurrent programs. 
When writing programs with concurrent threads accessing shared state, programmers -- however -- have to provide 
appropriate synchronisation among threads  to avoid access to inconsistent memory values. 
Software Transactional Memory (STM) (as proposed by Shavit and Touitou~\cite{Shavit1997})  
aims at providing programmers with an easily usable synchronisation technique 
for such an access to shared state.

STMs allow programmers to define software {\em transactions}, much alike data\-base transactions~\cite{Papadimitriou79b}. 
 A transaction consists of a number 
of read and write operations to the shared state, and the STM algorithm should guarantee 
these operations to take place ``seemingly atomically'', while ideally also allowing transactions to run concurrently. 
This seeming atomicity is formalized in a correctness criterion called {\em opacity}~\cite{Guerraoui2008}.
STMs typically try to avoid strict locking schemes in order to allow for good performance. 
This often comes at the price of complexity in verification as the high degree of concurrency 
leads to intricate interleavings. 
Moreover, opacity verification is a {\em parameterized verification} problem~\cite{DBLP:journals/ipl/AptK86}: 
STMs have to be proven correct for any number of transactions operating on any number 
of variables with moreover an unbounded number of possible values written to variables, 
i.e., for infinitely many possible instantiations.

A number of approaches have so far studied verification of STMs.  
The proposed techniques range
from model checking approaches of fixed instantiations~\cite{DBLP:conf/iceccs/BaekBKO10,DBLP:conf/icdcs/OLearyST09} 
over techniques employing data independence arguments to reduce the number of instantiation to look at~\cite{KW19} to
interactive proofs~\cite{Lesani2012,DBLP:conf/opodis/DohertyDDSW16,DBLP:journals/fac/DerrickDDSTW18}. 
Interactive approaches typically 
show refinement between the STM algorithm and an abstraction called TMS2 
which is known to be opaque~\cite{DGLM13}. 
Lesani~\cite{Lesani14} furthermore developed
a specific (also non-automatic) proof method for opacity by splitting opacity into a number
of other conditions (called {\em markability}).
In addition, Lesani and Palsberg~\cite{Lesani2013} have proposed conditions for {\em
disproving} opacity.  
A survey of verification approaches for STMs
can be found in~\cite{Lesani14,DBLP:conf/cost/CristalOCKKUTME15}. 

Interactive proofs provide results for the parameterized verification problem 
of a specific STM but are often very laborious, requiring several weeks of work in particular for defining invariants. 
Hence, it seems to be attractive to employ automatic model checking for all instantiations at once as developed by 
Guerraoui, Henzinger and Singh~\cite{GuerraouiHS10} (in the following referred to as the GHS
approach). This approach is based on a {\em small model theorem} allowing to reduce 
the parameterized verification problem to a model checking problem over 2 transactions 
and 2 variables. GHS applied their technique to a number of STMs, including DSTM~\cite{DBLP:conf/podc/HerlihyLMS03}
and TL2~\cite{DBLP:conf/wdag/DiceSS06}. Abdulla et al.~\cite{DBLP:conf/date/AbdullaDRSZ13} have 
furthermore employed the same technique for verification of a hybrid TM. 
Our interest was thus in the applicability of this small model theorem to 
STMs like the pessimistic STM in~\cite{DBLP:conf/opodis/DohertyDDSW16} or 
like FastLane~\cite{DBLP:conf/sefm/SchellhornWTKW18}, 
both of which have been interactively verified. 
This paper reports about the outcome of this investigation: 
We re-investigate the applicability of the small model theorem of GHS to 
opacity checking of software transactional memory algorithms.

\section{Background}

We start by explaining software transactional memory algorithms and the property of interest, {\em opacity}. 
In this, we follow GHS~\cite{GuerraouiHS10}, not the standard definition, because their small model theorem 
is given for their own, specific version of opacity. 
Later, we will comment on differences to the standard definition. 

\subsection{Basics} 

Software transactional memory algorithms allow for concurrent access to
shared state. 
The locations to be accessed by the STM are a set $\Var$ of variables. 
Programmers can use commands $C =\{\commit\} \cup (\{\rd, \wrt\} \times \Var)$  
(commit, read, write) to interact with STMs (plus typically an operation \texttt{begin} 
which is however not formalized by GHS).
The STM algorithm might respond to these commands by aborting a transaction, thus we let 
$\hat{C} = C \cup \{\abort\}$. 
We use $T=\{1, \ldots, n\}$ as the set of transaction or thread identifiers\footnote{Guerraoui et al.~distinguish between transactions and threads, 
but for reasons of simplicity we have refrained from doing so here.} and let 
$S = C \times T, \hat{S} = \hat{C} \times T$. 
We write such statements $((c,v),t)$ as $c_t(v)$ as e.g.~in $\rd_t(x)$ 
stating that thread $t$ reads from variable $x$.  
Note that Guerraoui et al.~do not consider the {\em values} (of variables) as 
passed as parameters to writes or returned as outputs of reads.

Opacity is defined by looking at the histories an STM algorithm produces. 
A {\em history} is a word $h \in \hat{S}^*$, i.e., a sequence of statements. 
The {\em projection} of a history $h$ on a transaction $t$,  
$h|_t$, consists of the statements in $h$ for transaction $t$ only.
We assume transaction identifiers to be unique. 
In a projection $h|_t = s_0 \ldots s_m$, either $s_m$ is an abort or commit statement 
or these statements do not occur in  $h|_t$ at all. In the latter case, the transaction is {\em live} in $h$, in the former 
it is {\em finished}. If the last statement is $\commit$, the transaction is {\em committing}; 
if it is $\abort$, the transaction is {\em aborting}. 

For two transactions $t_1, t_2$, we say that $t_1$ {\em precedes} $t_2$ in $h$, $t_1 <_h t_2$, 
if the last statement of $t_1$ occurs before the first statement of $t_2$. 
If neither $t_1 <_h t_2$ nor $t_2 <_h t_1$, then transactions $t_1$ and $t_2$ are 
{\em concurrent} in the history $h$. A history is {\em sequential} if no transactions are concurrent.

A transaction $t$ {\em writes} to a variable $x$ in a history $h$ if $h$ contains a statement 
$\wrt_t(x)$. A statement $s=\rd_t(x)$ in $h$ is a {\em global read} of variable $x$ if there is no $\wrt_t(x)$ before 
$s$ in $h$, i.e., $t$ does not read from its own write.  As an example, consider the sequential history 
\[ h_1  =  \wrt_1(x) \commit_1 \rd_2(x) \commit_2 \]

\noindent consisting of transactions 1 and 2. Both are committing transactions, $1$ precedes 2; 
transaction 2 has a global read of variable $x$, transaction 1 writes to $x$. 

Note that operations in histories are often split into invocations and responses in other formalizations of opacity, 
but in this we again stick to the formalization of GHS. 

\subsection{Opacity}

STM algorithms can broadly be categorized as using {\em direct} or {\em deferred updates}. 
In a direct update algorithm, the variables in shared state are directly updated during write operations; 
in deferred update algorithms the actual update takes place during commit operations. 
For the standard definition of opacity as given in~\cite{Guerraoui2008} this differentiation does not matter: 
the same definition of opacity is applicable to direct and deferred update algorithms. 
The definition of GHS is however based on {\em conflicts} between statements 
which allows to ignore the actual values of variables -- at the price of needing to introduce a definition of opacity differing 
from the original one. 

A statement $s_1$ of transaction $t_1$ is in {\em conflict} with a statement $s_2$ in $t_2$, $t_1 \neq t_2$, 
in a history $h$ if (i) $s_1$ is a global read of some variable $x$, $s_2$ is a $\commit$ statement 
and $t_2$ writes to $x$, or 
(ii) both $s_1$ and $s_2$ are $\commit$ statements and $t_1$ and $t_2$ both write to some variable $x$. 
Opacity requires ``seeming atomicity'' of transactions which is formalized by 
the existence of a sequential history reordering the concurrent one, 
however by keeping the real-time order and the order of conflicting statements. 
The following gives a definition of opacity for STMs with deferred update. 

\begin{definition}
 A history $h$ is {\em opaque} if there is a sequential history $h'$ such that 
 $h$ and $h'$ are strictly equivalent, i.e., 
\begin{itemize}
    \item for all transactions $t$: $h|_t = h'|_t$, 
   \item for all transactions $t_1, t_2$, if $t_1 <_h t_2$ and the last statement of 
    $t_1$ is a commit or abort, then $t_1 <_{h'} t_2$ and 
    \item for every pair of statements $s_i, s_j$ in $h$, if $s_i$ is in conflict with $s_j$ 
       and $i < j$, then $s_i$ occurs before $s_j$ in $h'$. 
\end{itemize}
\end{definition}

We refer to such a definition as a {\em conflict-based} definition of opacity. 
As GHS say, a corresponding definition can be given for STMs with direct update, 
but their paper does not contain it and also does not provide a small model theorem 
for STMs with direct update. 

As an example, consider again history $h_1  =  \wrt_1(x) \commit_1 \rd_2(x) \commit_2$. 
In $h_1$, statement $\rd_2(x)$ and $\commit_1$ conflict. 
The sequential history witnessing opacity of $h_1$ is $h_1$ itself. 
As an example of a non-opaque history consider $h_2$:
\[ h_2 = \wrt_1(x) \wrt_1(y) \rd_2(x) \commit_1 \wrt_2(y) \commit_2 \]

\noindent
Here, $\rd_2(x)$ and $\commit_1$ conflict as well as $\commit_1$ and $\commit_2$. 
Thus a sequential history strictly equivalent to $h_2$ would need to order transaction 
1 before 2 and 2 before 1 which is impossible.

\section{An Example STM: DSTM}

As an example STM, we use the DSTM algorithm (Dynamic STM) of Herlihy et al.~\cite{DBLP:conf/podc/HerlihyLMS03} 
in the version given by Lesani and Palsberg~\cite{Lesani2013}. 
This version is called CoreDSTM.

Though not formalized in the histories of conflict-based opacity definitions, STM algorithms write {\em values} 
into variables and read operations need to return such values. 
We let $\Val$ be the set of values for variables.  
CoreDSTM employs the following so called {\em meta data} to ensure opacity: 
\begin{itemize}
  \item $status : T \rightarrow \{\bbC,\bbA,\bbR\}$ (the status of every transaction), 
  \item $rdSet: T \rightarrow 2^{\Var \times \Val}$ (the variables read by a transaction), and 
  \item $state: \Var \rightarrow T \times \Val \times \Val$ (last writer with old and new value). 
\end{itemize}

\noindent Initially, $rdSet = \lambda t. \emptyset$, $status = \lambda t.\bbC$ and 
$state = \lambda x.(t_0,0,0)$ where $t_0$ is some dedicated transaction initially 
setting all variable values to 0. The component $state$ stores the last transaction having written to a variable as well as 
the old and new value. We access the three components of a state $st$ 
by $st.writer \in T$, $st.\mathit{newVal} \in \Val$ and $st.\mathit{oldVal }\in \Val$. The current status of transactions 
($\bbC$ = committed, $\bbA$ = aborted and $\bbR$ = running) is recorded in $status$. 
The STM furthermore tracks the set of variables read by transactions together with the value read.

\begin{algorithm}
\caption{CoreDSTM}\label{alg:dstm}
\begin{varwidth}[t]{0.54\textwidth}
\begin{algorithmic}[1]
\Statex
\Procedure{read$_t$}{x}
\State s:= status(t);
\If {(s = $\bbA$)} 
    \State \Return $\bbA$;
\EndIf
\State st:= state(x);
\State v:= stableValue$_t$(st);
\State wr:= st.writer;
\If {(wr $\neq$ t)} 
   \State rdSet(t).add((x,v));
\EndIf 
\State valid := validate$_t$();
\If {($\neg$valid)}
    \State \Return $\bbA$;
\EndIf 
\State \Return v; 
\EndProcedure
\Statex
\Procedure{commit$_t$}{}
\State valid:= validate$_t$();
\If {($\neg$valid)}
   \State \Return $\bbA$;
\EndIf
\State b:= status(t).CAS($\bbR$,$\bbC$);
\If {(b)} 
   \State \Return $\bbC$;
\Else
   \State \Return $\bbA$;
\EndIf 
\EndProcedure
\Statex
\Procedure{stableValue$_t$}{st}
\State t´:= st.writer;
\State s´:=status(t´);
\If {(t´ $\neq$ t $\wedge$ s´ = $\bbR$)} 
   \State status(t´).CAS($\bbR, \bbA)$;
\EndIf
\State s´´ = status(t´);
\If {(s´´=$\bbA$)}
   \State v:= st.oldVal;
\Else
   \State v:= st.newVal;
\EndIf
\State \Return v; 
\EndProcedure
\end{algorithmic}
\end{varwidth}
\hfill 
\begin{varwidth}[t]{0.54\textwidth}
\begin{algorithmic}[1]
\Statex
\Procedure{write$_t$}{x,v}
\State s:=status(t);
\If {(s = $\bbA$)} 
    \State \Return $\bbA$;
\EndIf
\State st:= state(x);
\State wr:= st.writer;
\If {(wr=t)}
   \State st.newVal := v;
   \State \Return ok;
\EndIf
\State v´ := stableValue$_t$(st);
\State st´:= (t,v´,v);
\State b:= state(x).CAS(st,st´);
\If {(b)}
   \State \Return ok;
\Else
   \State \Return $\bbA$;
\EndIf
\EndProcedure
\Statex
\Procedure{validate$_t$}{}
   \ForAll {((x,v) $\in$ rdSet(t))}
       \State st:= state(x);
       \State t´:=st.writer;
       \State s´:=status(t´);
       \If {(s´=$\bbC$)}
          \State v´:=st.newVal;
       \Else
          \State v´:=st.oldVal;
       \EndIf
       \If {(v $\neq$ v´)}
            \State \Return false;
       \EndIf
   \EndFor
   \State s:=status(t);
   \State \Return (s=$\bbR$);
\EndProcedure
\end{algorithmic}
\end{varwidth}
\end{algorithm}

Algorithm~\ref{alg:dstm} gives the code for read, write and commit operations. 
The operation CAS used therein has the following meaning: in a statement \texttt{var.CAS(o,n)}  
the value of \texttt{var} is compared to \texttt{o} (old) and -- if equal -- is set to \texttt{n} (new). 
This compare-and-set is done in one {\em atomic} step. The CAS operation returns the result 
of the comparison, i.e., a boolean operation. 
We see that the write operation first of all stores the value to be written in \texttt{newVal} 
(line 8 within write). Procedures \texttt{validate} and \texttt{stableValue} 
only retrieve the new value if the writing transaction has committed. 
Hence, this is a deferred update algorithm. 

As observed by Lesani and Palsberg, CoreDSTM is not opaque. 
The actual implementation however seems to differ from this version and is opaque.
For demonstration purposes here it makes sense to look at the non-opaque version. 
The non-opacity of CoreDSTM can be seen in the following execution:
Transactions 1 and 2 first both read from variables $x$ and $y$ (the initial value 0).
Afterwards transaction 1 writes to $x$ (say, value 7) and transaction 2 to $y$ (say, value 8).
Then they commit concurrently: first, transaction 1 executes lines 15 and 16 of 
commit, then transaction 2  does so
(both having their local variable \texttt{valid} being true afterwards), 
and after that they successfully end their commit operation. 
As a history\footnote{For determining the history with atomic operations, we in principle need to 
fix the linearization point of the operation. Whatever this might be for the commit, all choices 
lead to the same history.}, this gives 
\begin{align}
h = \rd_1(x) \rd_2(x) \rd_1(y) \rd_2(y) \wrt_1(x) \wrt_2(y) \commit_1 \commit_2 \label{his}
\end{align} 

This history is not opaque as $\rd_2(x)$ and $\commit_1$ as well as $\rd_1(y)$ and $\commit_2$ 
are in conflict, so the required ordering for the sequential history is that 2 has to precede 1 
and 1 has to precede 2 which cannot be fulfilled at the same time.

\section{The Small Model Theorem} 

GHS aim at an automatic way of checking opacity (as well as strict serializability). 
To this end, they develop four properties of STMs which are sufficient for reducing the general verification 
problem to a model checking problem over 2 variables and 2 transactions\footnote{In their 
formalization, 2 threads.}. 
We (informally) introduce these four properties here, and study one of them in more detail later.

 The properties refer to the executions of a particular STM algorithm $M$ as 
seen in its histories.  

\begin{description}
  \item[P1] Transaction projection: Let $h$ be a history of an STM $M$ and $T' \subseteq T$ 
     the set of all committed plus some of the live transactions of $h$. Then $h|_{T'}$ is a 
     history of $M$ as well. 
  \item[P2] Thread symmetry: Plays no role in our formalization as we do not distinguish between threads 
     and transactions. 
  \item[P3] Variable projection: Let $h$ be a history of an STM $M$ without aborting transactions, 
      and let $V \subseteq \Var$. Then $h|_V$ is a history of $M$ as well (where the projection of 
     $h$ onto some set of variables removes all reads and writes to other variables). 
  \item[P4] Monotonicity\footnote{This is the version of property P4 for monotonicity taken from the PhD thesis of 
       one of the authors~\cite{singh} to align it with the discussion on further conditions in~\cite{singh}.}: Let $h \cdot s$ be a history of an STM such that $h$ (a history) is opaque, $s$ (a single statement) 
     is not an abort statement, $h$ has exactly one live transaction and $s$ is a 
     statement of this transaction. Then there is some $h'$ which is  
     strictly equivalent to $h$ and  sequential,  
     and $h' \cdot s$ is a history of $M$. 
\end{description}

\noindent Here, $\cdot$ is concatenation. We will see below that when evaluated for concrete STMs property P4 is subject 
to interpretation. Intuitively, P4 states that whenever a history is allowed by an STM, then more sequential 
versions of the history are allowed as well. 

These four properties allow to reduce opacity checking of STMs to 2 transactions and 2 variables. 
An STM is said to be $(n,k)$-opaque if all histories with $n$ transactions (i.e., at most $n$ concurrent transactions) 
and $k$ variables are opaque. 

\begin{theorem}
If a TM $M$ ensures (2,2)-opacity and 
satisfies the properties {\bf P1, P2, P3} and {\bf P4} for opacity,
then $M$ ensures opacity.
\end{theorem} 

\noindent For the proof, see \cite{singh}. The proof proceeds by constructing for every non-opaque 
history of $M$ another non-opaque history with just 2 variables and 2 transactions. 
Properties P1 to P4 ensure that this new history is still possible for $M$. Thus every violation of 
opacity can be seen in histories with 2 transactions and 2 variables. 
Hence automatic model checking of STMs is possible by inspecting instantiations with 2 transactions and 
2 variables only. 

\section{Applicability}

There are a number of issues making this small model theorem and its associated automatic model checking procedure difficult to apply 
to concrete STM algorithms. 

\begin{observation}
  It is unclear how to automatically show 
  properties  P1, P2, P3 and P4 for some 
  concrete STM algorithm. 
\end{observation} 

Given that the technique is supposed to make opacity model checkable, this is a 
realistic difficulty. The PhD thesis of Singh~\cite{singh} employs three other conditions 
(abort isolation, pending isolation and conflict commutativity) to guarantee P1 and P4. 
We exemplify Issue 1 on one of them, namely abort isolation. 

\begin{definition}
    A TM algorithm is {\em abort isolated} if for every history $h$ and every aborted 
     transaction $t$ in $h$, the following holds: if an instruction of $t$ changes the value of 
     a global variable $g$ and a transaction $t'$ observes the value of $g$ before $t$ aborts, 
     then $t'$ aborts in the step of observing $g$. 
\end{definition}

Abort isolation together with a similar condition for live transactions guarantees property P1~\cite{singh}. 
Singh writes that DSTM is abort isolated because ``an aborted transaction does not change 
the state in DSTM''. An automatic way of showing abort isolation is not proposed. 
However, CoreDSTM (as well as DSTM) can produce histories of the following form: First, a transaction $t$ writes to a 
variable (thus setting the writer of this variable to $t$ thereby changing the value of a global 
variable). Afterwards a further transaction $t'$ writing to the same variable would see $t$ as writer and 
abort $t$.  Transaction $t'$ might successfully commit later. So DSTM is not abort isolated. 

\begin{observation}
The opacity definition (and hence the theorem) refers to STMs with deferred update only. 
\end{observation} 

It is not clear whether such a reduction theorem also holds for STMs with direct update and 
what the exact formulation of properties P1 to P4 would be in that case. 
Moreover, not all STMs strictly fall in one or the other category. 
An example for this is the STM FastLane~\cite{DBLP:conf/ppopp/WamhoffFFRM13}
which provides two different modes for transactions: one master transaction 
uses direct update while all helper transactions employ deferred update. 
FastLane has been shown to be opaque using interactive theorem proving~\cite{DBLP:conf/sefm/SchellhornWTKW18}.

The next issue refers to the definition of opacity. 
The standard reference for the definition of opacity (as also given by GHS) is that of 
Guerraoui and Kapalka~\cite{Guerraoui2008}. 
There are two key differences between the definition given there and the one employed for the small model theorem:  
\begin{enumerate}
  \item Transactions operate on shared variables and these {\em possess a state}, 
     i.e., there are values associated with variables and these values appear in 
    histories as arguments or return values of write and read operations, 
  \item operations are divided into {\em invocations} and {\em responses} 
   (i.e., instead of a $\commit$ operation there are operations $\mathit{inv}(\commit,...)$ and 
    $\mathit{res}(\commit, ...)$).  
\end{enumerate} 

As a consequence of the first difference, opacity can and is then defined by looking at 
the values returned by reads (instead of by looking at conflicts) and by defining when 
these values are {\em legal} (namely when the last committing writer before a read has written this value). 
As a consequence of the second difference, histories can then directly describe {\em interleavings} 
of transactional operations (e.g., a commit of one transaction occurring concurrently with a commit of 
another transaction).  Both of these differences have consequence for the applicability 
of the small model theorem to model checking opacity. The following observation 
has already been made by us before~\cite{Koenig2017}.

\begin{observation}
The value-based definition of opacity is not the same as the conflict-based definition. 
\end{observation}

These two notions are in fact incomparable. 
The following two histories show that a value-based and a conflict-based definition 
of opacity do not coincide. For this, we extend the operations write and read with 
arguments and return values, respectively. That is, an operation $\wrt_1(x,7)$ is a 
write of transaction 1 on shared variable $x$ with value 7. 
\begin{eqnarray*}
  h_3 & = & \wrt_1(x,7) \commit_1 \rd_2(x,3) \commit_2 \\
  h_4 & = & \wrt_1(x,5) \wrt_2(x,5) \wrt_1(y,42)  \wrt_2(y,43) \commit_1 \rd_3(x,5) \commit_2 \rd_3(y,43) \commit_3 
\end{eqnarray*} 

\noindent History $h_3$ is opaque under the  conflict-based definition: statements $\rd_2(x,3)$ 
and $\commit_1$ are in conflict, thus transactions $1$ and $2$ need to be ordered as $1 < 2$ 
in the sequential history which is possible without violating other constraints on orderings. 
In a value-based definition of opacity $h_3$ is clearly not opaque since transaction 2 is reading 
an incorrect value from variable $x$.

On the other hand, under a value-based version history $h_4$ is opaque (as justified by the 
sequential order $1 < 2 < 3$). For the conflicts, we however get constraint $2 < 3$ 
(since $\commit_2$ and $\rd_3(y,43)$ are in conflict) as well $3 < 2$ (since $\rd_3(x,5)$ 
and $\commit_2$ are in conflict) 
which cannot both be satisfied by a sequential history. 

As a follow-up of moving to a conflict-based definition, GHS had to change 
the STM algorithms they employ as examples (as these  
 store values of shared variables, modify them by writes and 
return their values during reads).  For instance, they also give the DSTM algorithm (in the repaired version), 
but instead of recording old and new values of variables and using the stability check, 
they introduce an ownership set. 
The stated proof of opacity of DSTM thus refers to an abstraction of DSTM only.

The next  issue is the level of atomicity considered in the formalization of the GHS approach
and thus concerns the second difference in the opacity definition. 
First, because operations are not split into invocations and responses, all operations 
seem to be considered to be atomic. On a more detailed look\footnote{Guerraoui et al.~give the 
TM algorithms in a very unusual form. This makes it difficult to determine what the actual runs  
of a TM are, and what the histories derived from these runs are.}, 
TM algorithms considered by GHS still allow for runs interleaving statements of operations (via extended commands 
and specific $\bot$ responses). 
As this is invisible in the histories, the notion of ``a history being sequential'' gets 
unclear and as a consequence the interpretation of property P4 is unclear.

\begin{observation}
  It is unclear on which level of granularity property P4 is to be interpreted. 
\end{observation} 

\noindent For this, consider again history $h$ of CoreDSTM given in (\ref{his}). History $h$ is not opaque, but 
\[ h' = \rd_1(x) \rd_2(x) \rd_1(y) \rd_2(y) \wrt_1(x) \wrt_2(y) \commit_1 \]

\noindent is. With $h$ and $h'$ we have the situation required for property P4: $h' \cdot s$ where $s= \commit_2$ 
is not opaque, $s$ is not an abort, $h'$ is opaque.  
Property P4 states that the sequentialisation of $h'$ witnessing opacity of $h'$ 
which is 
\[ h'' = \rd_2(x) \rd_2(y) \wrt_2(y) \rd_1(x)  \rd_1(y)  \wrt_1(x)  \commit_1 \]
and its extension by $s$ needs to be a history of the STM. 
The question is now what the execution of the STM is which we need to look at 
when trying to get this history. 
The natural expectation is that $h''$ corresponds to an execution of 
DSTM in which {\em no} statements of $\commit_2$ are executed. 
However, then $h'' \cdot s$ is not a history of DSTM (transaction 2 will abort when 
it starts its commit after 1's commit) and consequently P4 is not satisfied.  
If we allow $h'' \cdot s$ to belong to an execution interleaving statements of the 
two commit operations, $h'' \cdot s$ is a history of DSTM (and P4 would be fulfilled). 
From the way property P4 is stated, such an $h''$ would however intuitively not be allowed 
as these histories are supposed to be ``sequential'' (which one can expect to be interpreted as 
``no interleaving of operations at all'').

The final observation also refers to the level of granularity in operations 
and hence possible interleavings of concurrent transactional operations.  

\begin{observation}
 The level of granularity of STM operations given by GHS differs to the published algorithms. 
\end{observation}

For example, for the (correct version of) DSTM as given by GHS, the read operation is an 
atomic step, write consists of two atomic steps (\texttt{write} and \texttt{own}) and 
commit of two steps (\texttt{validate} and \texttt{commit}, the \texttt{validate} for instance 
aborting several other transactions in a single atomic step).  
This simplifies in particular the proof of property P4 for DSTM, e.g., Singh~\cite{singh} 
argues that DSTM is conflict commutative (which guarantees P4) by saying 
``As the read consists of a single instruction, it cannot be concurrent with a commit instruction''.  
This clearly does not hold for CoreDSTM. 

STM implementations typically use specific instructions to achieve  
atomicity (like the CAS in DSTM) and use them only on a very fine-grained level as to allow  
a high degree  of concurrency. 
By employing  formal models of STMs extending atomicity to blocks of statements (as GHS do), 
critical interleavings leading to non-opaque 
behaviour can easily be missed.

\section{Conclusion}

In this paper, we have re-investigated the applicability of a small model theorem to 
the model checking of STMs with respect to opacity. 
While neither the small model theorem is incorrect nor the model checking results 
for the algorithms {\em as given in the paper} are, the general applicability of this theorem 
remains unclear. The usage of this theorem presupposes a number of 
abstractions carried out on (a) the opacity definition itself and (b) the STM 
algorithms being checked. The results employed by this technique are thus not 
directly transferable to the original STM algorithms.

\end{document}